\documentclass[aps,prl,twocolumn,superscriptaddress,floatfix,letter,longbibliography]{revtex4-2}

\usepackage{epsfig,amsmath,amssymb,color,amsfonts,physics,microtype}
\usepackage[bookmarks=true,colorlinks,citecolor=red]{hyperref}
\usepackage{float}
\usepackage{dsfont}
\usepackage{wrapfig}
\usepackage{tikz}
\usepackage{physics}
\usepackage{mathrsfs}
\usepackage[export]{adjustbox}
\usetikzlibrary{quantikz}
\usepackage{soul}
\usepackage{comment}
\usepackage{orcidlink}

\newcommand{\Rey}{R}
\newcommand{\Sc}{S}

\begin{document}
\title{Magic of the Well: assessing quantum resources of fluid dynamics data}

\author{Antonio Francesco Mello~\orcidlink{0009-0002-4401-1126}}
\affiliation{
Center for Computational Quantum Physics, Flatiron Institute, New York, NY 10010, USA
}
\affiliation{International School for Advanced Studies (SISSA), via Bonomea 265, 34136 Trieste, Italy}

\author{Mario Collura~\orcidlink{0000-0003-2615-8140}}
\affiliation{International School for Advanced Studies (SISSA), via Bonomea 265, 34136 Trieste, Italy}
\affiliation{INFN, Sezione di Trieste, Via Valerio 2, 34127 Trieste, Italy}

\author{E. Miles Stoudenmire}
\affiliation{
Center for Computational Quantum Physics, Flatiron Institute, New York, NY 10010, USA
}

\author{Ryan Levy~\orcidlink{0000-0003-4952-7156}}\thanks{Current affiliation: PsiQuantum, 700 Hansen Way, Palo Alto, CA 94304, USA } 
\affiliation{
Center for Computational Quantum Physics, Flatiron Institute, New York, NY 10010, USA
}
\date{\today}

\begin{abstract}
    We investigate the quantum resource requirements of a dataset generated from simulations of two-dimensional, periodic, incompressible shear flow, aimed at training machine learning models. By measuring entanglement and non-stabilizerness on MPS-encoded functions, we estimate the computational complexity encountered by a stabilizer or a tensor network solver applied to Computational Fluid Dynamics (CFD) simulations across different flow regimes. Our analysis reveals that, under specific initial conditions, the shear width identifies a transition between resource-efficient and resource-intensive regimes for non-trivial evolution. Furthermore, we find that the two resources qualitatively track each other in time, and that the mesh resolution along with the sign structure play a crucial role in determining the resource content of the encoded state.
    These findings offer useful guidelines for the development of scalable, quantum-inspired approaches to fluid dynamics. 
 \end{abstract}

\maketitle
\paragraph{Introduction.---}
The Navier-Stokes equations, whose closed-form solutions are known only in specific limits, represent one of the most challenging problems in physics and mathematics~\cite{fefferman2006existence}. Computational fluid dynamics (CFD) methods have been devised over the years to numerically address these equations, with many successful results that have significantly advanced our understanding of nature~\cite{orszag1972numerical, kolmogorov1941local}. However, most of these techniques are affected by the so-called \textit{curse of dimensionality}. Indeed, the pressure and velocity fields are represented as multi-dimensional arrays whose dimension depends on a mesh associated to the discretization of the spatial domain. As a matter of fact, the precise description of some kinds of turbulence requires the size of the mesh to grow exponentially with the Reynolds number (\Rey), making the problem beyond reach for mesh-based solvers~\cite{popeturbulence}. 

Quantum systems exhibit an analogous exponential complexity, which was first understood to stem from their entanglement structure~\cite{eisert2010colloquium, Calabrese_2004, preskill2012quantum}. In the quantum realm, tensor network (TN) methods have established themselves as powerful tools to study the equilibrium and out-of-equilibrium properties of low-entangled one-dimensional systems, namely using the matrix product state (MPS) formalism~\cite{Vidal_2003, Schollwck2011,PAECKEL2019167998, Orus2014annphys}. Recently, the same TN compression principles have been applied to high-dimensional classical problems in addition to quantum systems. This has paved the way for the field of quantum-inspired CFD solvers~\cite{pisoni2025compressionsimulationsynthesisturbulent, gourianov2025tensor, gourianov2022quantum, Ye_Loureiro_2024,Ye2022poisson,ortega2025tensor,danis2025tensor,peddinti2025technicalreportquantuminspiredsolver, gómezlozada2025simulatingquantumturbulencematrix}, where the MPS encoding is leveraged to solve fluid dynamics equations bypassing the exponential dense vector representation. Nevertheless, it is well known that entanglement is not the only resource determining quantum complexity. Remarkably, an equally essential quantum resource is represented by non-stabilizerness~\cite{gottesman1997stabilizer, gottesman1998heisenberg, aaronson2004improved}, which characterizes the extent to which a quantum protocol deviates from classical polynomial simulability as defined by the Gottesman-Knill theorem~\cite{gross2006hudson, Gross2021, bravyi2005UniversalQuantumComputation, bravyi2019simulation}. Within this framework, highly entangled stabilizer states can be efficiently simulated with classical resources. 

Understanding the interplay between entanglement and non-stabilizerness remains an open problem, and a number of protocols have been proposed that leverage the strengths of both~\cite{mello2024ctdvp, PhysRevLett.133.150604, lami2024quantum, qin24cdmrg,qian2024augmentingfinitetemperaturetensor,qin25ctdvp, mello2025cliffordloschmidt,huang2024cliffordcircuitsaugmentedmatrix}. Recently, the ability of neural quantum states (NQS) to encode both resources has been also investigated~\cite{sinibaldi2025nonstabilizernessneuralquantumstates}, as well as possible extensions to fermionic gaussian systems~\cite{collura2025quantummagicfermionicgaussian}.
\begin{figure}[t]
\includegraphics[width=0.5\textwidth]{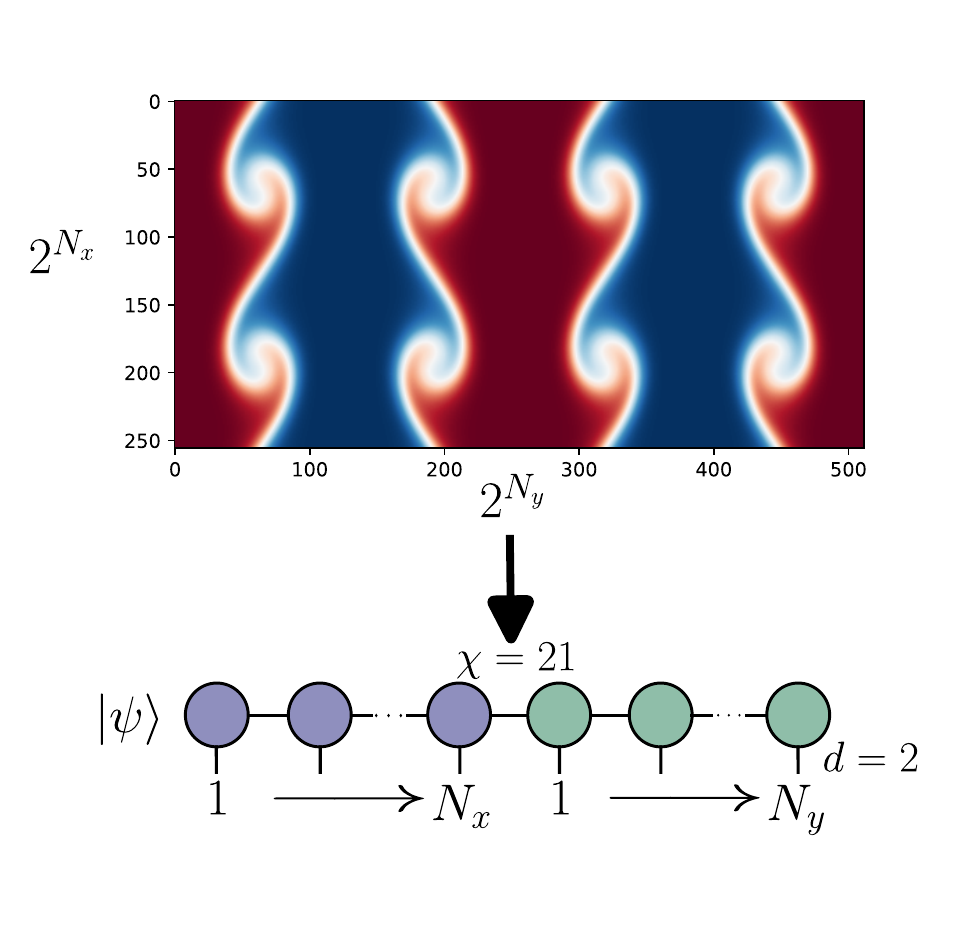}
\caption{Schematic representation of the tracer field $s_{ij}$ (\textit{top}), which is then encoded as an MPS (\textit{bottom}). The encoding of the shown example has a bond dimension $\chi=21$, which is $\approx 8\%$ of the possible maximum $\chi_{max}$. 
The tracer field is obtained as a solution of eq.~\eqref{eq:pde_sf}.}
\label{fig:sketch}
\end{figure}
In this work, we evaluate the quantum resource content of a dataset representing shear flow. The dataset, generated for training machine learning algorithms, is derived from simulations of a two-dimensional, periodic, incompressible shear flow -- a fluid regime where adjacent layers continuously deform as they slide past one another at different velocities. Shear flow is a phenomenon observed in a variety of scenarios, ranging from rivers~\cite{koyama2008numerical,olsen1995three} and atmospheric boundary layers~\cite{khanna1998three} to industrial processes involving transport of fluids~\cite{hanspal2006numerical, hosain2018fluid}. Its evolution is described by two partial differential equations: one for the pressure and velocity fields, and one for the tracer which is obtained from the velocity and allows the shear to be visualized.
Specifically, we encode the solutions for the tracer and pressure fields as MPS, and estimate the complexity faced by stabilizer or TN CFD solvers. This is accomplished by assessing their entanglement and non-stabilizerness across various flow regimes. We measure entanglement through the von Neumann entanglement entropy, while, among the several measures of non-stabilizerness~\cite{Beverland2020, howard2017robustness, heinrich2019robustness}, we adopt stabilizer Rényi entropies (SREs)~\cite{leone2022stabilizer} for the ease with which they can be evaluated numerically~\cite{tarabunga2023critical, tarabunga2023manybody, tarabunga2024magic}, and in experiments~\cite{niroula2023phase}. Varying Reynolds (\Rey) and Schmidt (\Sc) numbers across a range of values, we find that the two considered resources display an analogous qualitative behavior, reflecting the intrinsic complexity of the problem. With the most oscillatory initial conditions of the dataset, the shear width delineates a boundary between a resource-intensive regime and a resource-efficient one. Finally, we highlight how the grid resolution and the sign structure influence the quantum resource content of the encoded state. 
Our analysis reveals that, for the chosen dataset, non-stabilizerness and entanglement shape the resource cost of shear-flow simulations in a similar manner, and provides a framework to identify regimes where stabilizer-TN solvers could outperform classical methods.\\

\begin{figure}[h!]
\includegraphics[width=0.45\textwidth]{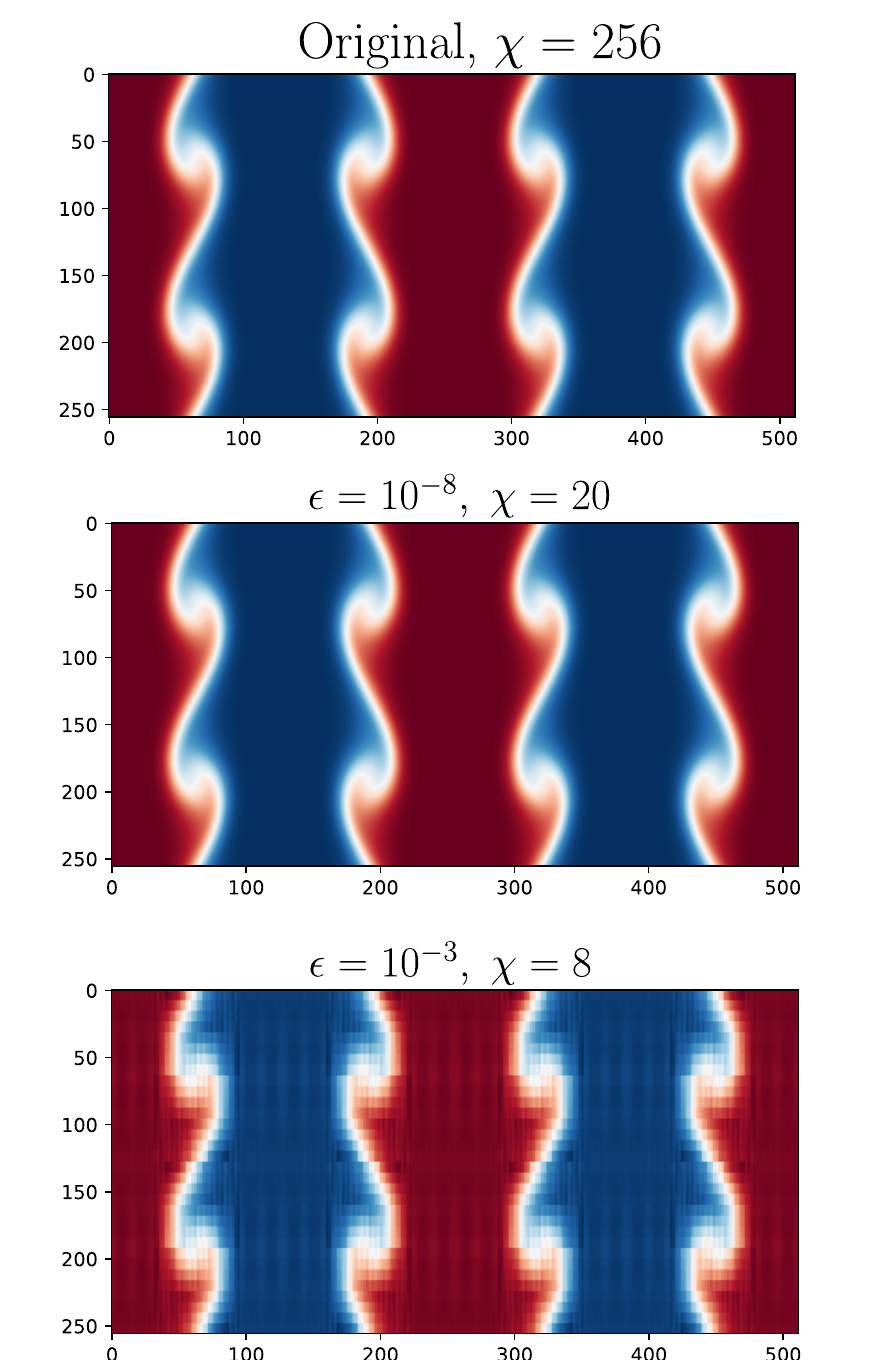}
\caption{Visual representation of the encoded data for the tracer field. From top to bottom: original data in~\cite{ohana2024well}; image obtained after a lossless MPS encoding (accuracy w.r.t. the original image, as measured through the root mean squared error, is $\approx 99.7 \%$); image obtained after high compression of the MPS encoding (fine details are lost).}
\label{fig:compression}
\end{figure}

\begin{figure*}[t]
\includegraphics[width=\textwidth]{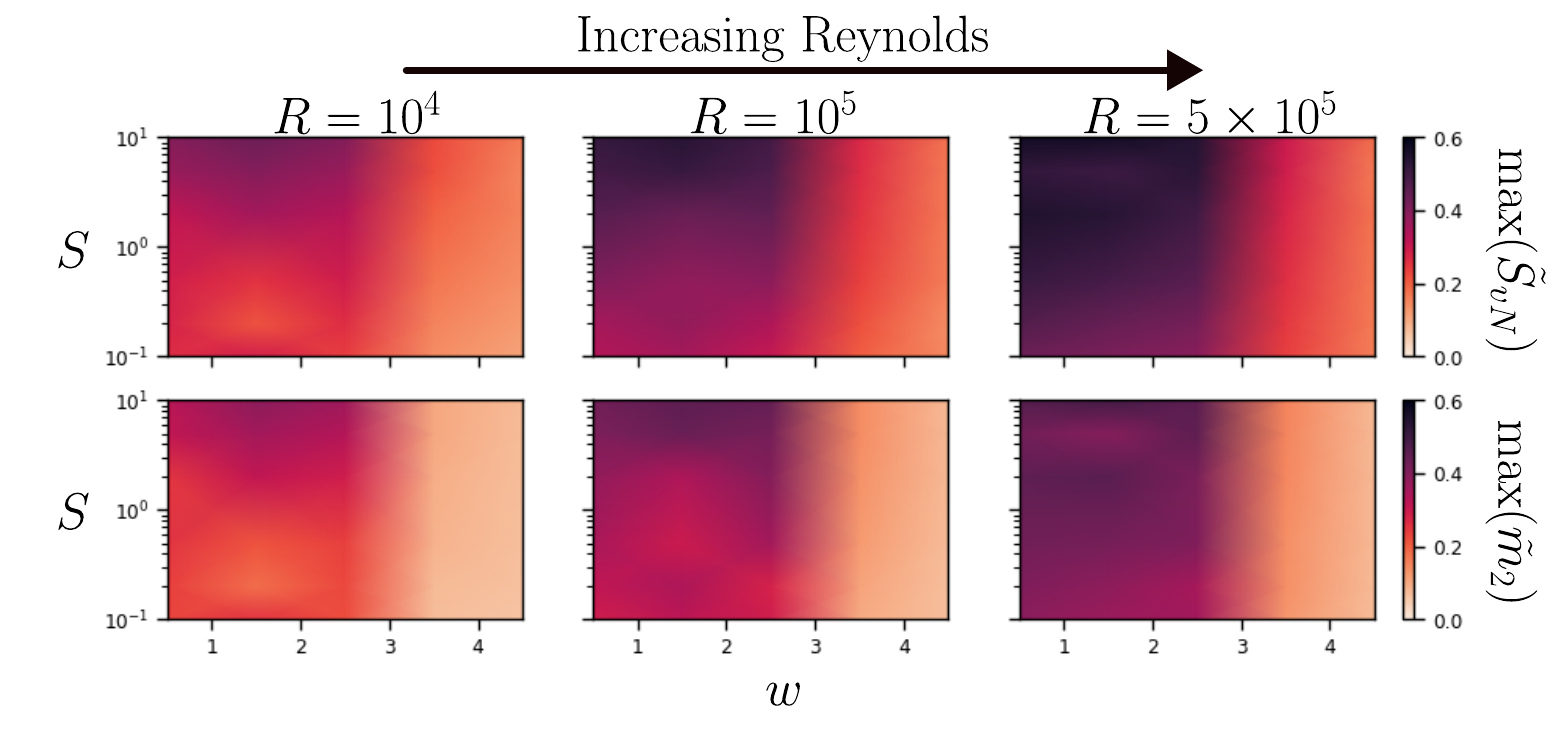}
\caption{Resource phase diagrams of the shear flow data illustrating how entangled (\textit{bottom row}) or magic (\textit{top row}) content each simulation has. Each column represents a different Reynolds number with the top row showing maximum normalized entanglement entropy $\tilde{S}_{vN}$ and the bottom row showing maximum normalized magic content $\tilde{m}_2$. Each plot shows the Schmidt number $S$ as a function of initial shear width factor $w$. The shear width factor is inversely proportional to how sharply the transition between the two flows are. Interpolation was used between data points. }
\label{fig:phases}
\end{figure*}

\paragraph{Non-stabilizerness preliminaries.---}
Since we encode the classical functions as MPS with local dimension $2$ (see Sec.~\textit{Methods}), we introduce the basics of non-stabilizerness by considering a quantum system composed of $N$ qubits. The set of Pauli matrices is defined as $\{\hat{\sigma}^{\mu} \}_{\mu=0}^3$, where $\hat{\sigma}^0 = \hat{I}$ is the identity operator. The local computational basis $\{ \ket{0}, \ket{1} \}$  corresponds to the eigenstates of $\hat{\sigma}^3$ , such that $\hat{\sigma}^3 \ket{\sigma} = (-1)^\sigma \ket{\sigma},\ \text{for } \sigma = 0, 1$.
A general Pauli string $\hat{P}$ is specified by a multi-index $\boldsymbol{\mu} = \{ \mu_1, \dots, \mu_N \}$, where $\mu_j \in \{0,1,2,3\}$. The corresponding operator is $\hat{P}(\boldsymbol{\mu}) = \bigotimes_{j=1}^N \hat{\sigma}^{\mu_j}_j$, and the set $\mathcal{P}_N $ contains all such Pauli strings, excluding global phases. The full $ N $-qubit Pauli group is defined as $\tilde{\mathcal{P}}_N = \{ \pm 1, \pm i \} \times \mathcal{P}_N$. A crucial class of quantum operations is the Clifford group. Clifford unitaries are defined as those that map Pauli strings to other Pauli strings under conjugation. Formally, the Clifford group is the normalizer of the Pauli group and is defined as $\mathcal{C}_N = \left\{ \hat{U} \, \middle| \, \hat{U} \hat{P} \hat{U}^\dagger \in \tilde{\mathcal{P}}_N \text{ for all } \hat{P} \in \tilde{\mathcal{P}}_N \right\}.$
This group can be generated by the Hadamard, phase, and CNOT gates.

Given these ingredients, we define \emph{stabilizer states} as the set of quantum states that can be constructed from the initial product state $\ket{0}^{\otimes N}$ using only Clifford operations. Despite potentially generating high entanglement, these states can be simulated efficiently on classical computers, meaning they do not offer a quantum computational advantage.

To quantify non-Clifford resources, we introduce a measure of deviation from stabilizer states, often referred to as \emph{magic} or \emph{non-stabilizerness}. Stabilizer Rényi entropies stand out as well-suited metric for this resource. For a pure quantum state $\ket{\Psi} $, the SRE of order \( \alpha \) is defined as
\begin{equation}
M_\alpha = \frac{1}{1 - \alpha} \log \left[ \frac{1}{2^N} \sum_{\hat{P} \in \mathcal{P}_N} \left( \frac{ \langle \Psi | \hat{P} | \Psi \rangle }{ \langle \Psi | \Psi \rangle } \right)^{2\alpha} \right]
\end{equation}
where  $\alpha$ is the Rényi index.
Defining the probability distribution over $ \mathcal{P}_N $ as
$\displaystyle \Xi_{\hat{P}} = \frac{1}{2^N} \left( \frac{ \langle \Psi | \hat{P} | \Psi \rangle }{ \langle \Psi | \Psi \rangle } \right)^2,$
the SREs can be expressed as averages with respect to  $\Xi_{\hat{P}}$:
\begin{subequations}
\label{eq:pauli_sampling}
\begin{align}
M_\alpha &= \frac{1}{1 - \alpha} \log \mathbb{E}_{\hat{P} \sim \Xi_{\hat{P}}} \left[ \Xi_{\hat{P}}^{\alpha - 1} \right] - N \log 2, \quad \text{for } \alpha \neq 1, \\
M_1 &= - \mathbb{E}_{\hat{P} \sim \Xi_{\hat{P}}} \left[ \log \Xi_{\hat{P}} \right] - N \log 2.
\end{align}
\end{subequations}

For pure states of qubits, it has been shown that SREs with $\alpha \geq 2$ are monotonic under stabilizer operations and thus valid as resource monotones~\cite{leone2024stabilizer, haug2023stabilizer}. In the following, we focus on the case $\alpha = 2$.
When the many-body state $\ket{\Psi}$ admits an efficient MPS representation, $M_\alpha$ can be efficiently evaluated by directly sampling from $\Xi_{\hat{P}}$~\cite{Lami2023} or by writing it in the Pauli basis~\cite{tarabunga2024nonstabilizerness}.\\

\begin{figure*}[t]
\includegraphics[width=\textwidth, inner]{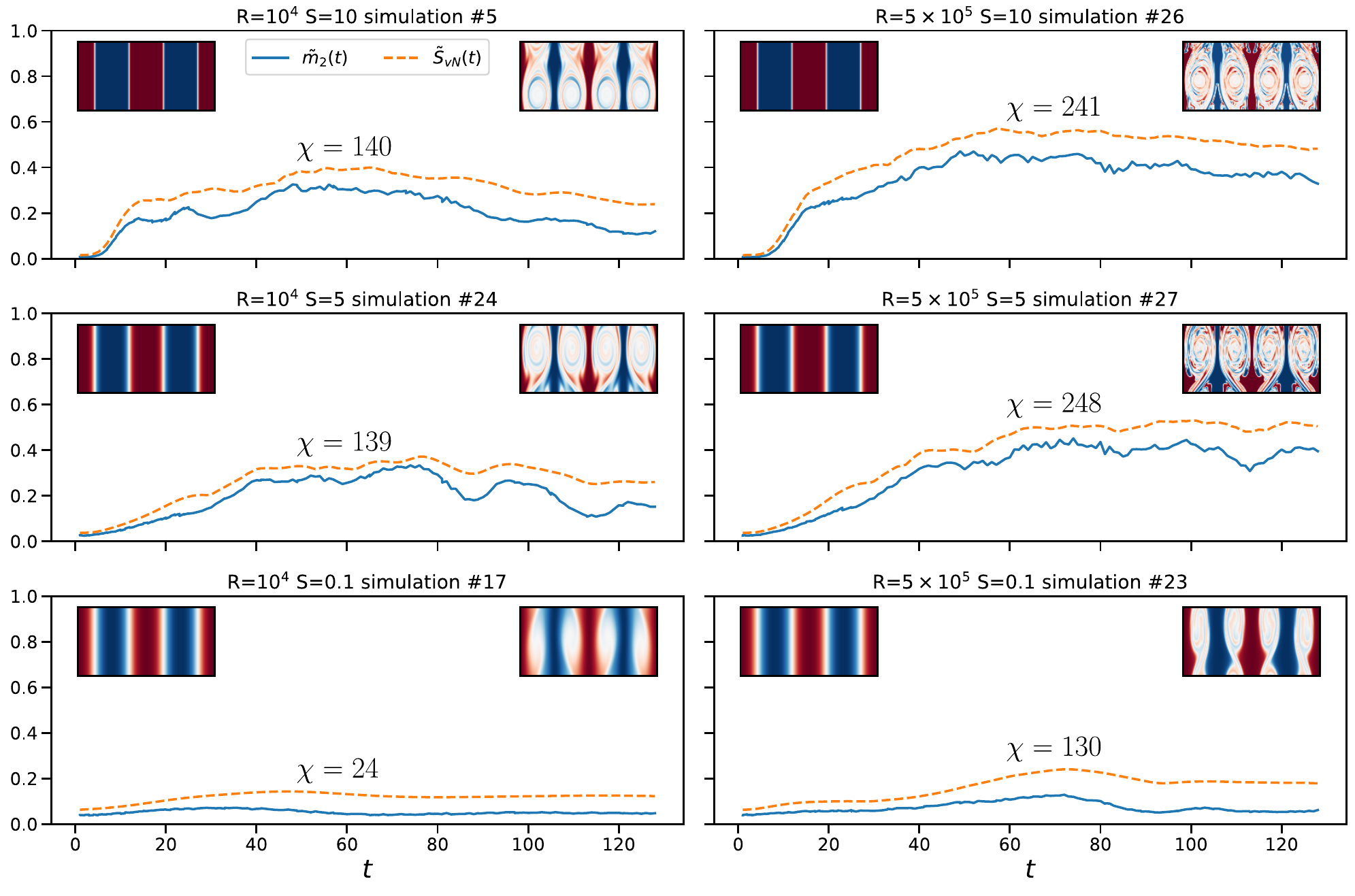}
\caption{Comparison of the normalized non-stabilizerness $\tilde{m}_2(t)$ and normalized entanglement entropy $\tilde{S}_{vN}(t)$ for the tracer field as a function of time for different values of $R$, $S$, and initial shear width $w \in [1,3,4]$. The insets highlight the initial and final stages of the evolution, while $\chi$ indicates the maximum bond dimension attained in each case for the MPS encoding. The global truncation error for $t=128$ of each simulation is shown in Fig.~\ref{fig:all_trunc_err} in the appendix.} 
\label{fig:m_svn_comp}
\end{figure*}

\paragraph{The model.---}
In the following, we make use of the generated data of Ref.~\cite{ohana2024well}. Specifically, we consider a two-dimensional periodic shear flow governed by the incompressible Navier-Stokes equations
\begin{equation}\label{eq:pde_sf}
\begin{cases}
    \displaystyle
\frac{\partial u }{\partial t} + \nabla p -\nu \Delta u = -u \cdot \nabla u \\ 
    \displaystyle
\frac{\partial s}{\partial t} -D\Delta s = -u \cdot \nabla s
\end{cases}
\end{equation}
where $\Delta = \nabla \cdot \nabla$ is the spatial Laplacian, $u=(u_x,u_y)$ is the velocity field, $s$ is the tracer and $p$ is the pressure satisfying the gauge condition $\int p = 0$. The parameters $D$ and $\nu$ represent diffusivity and viscosity, respectively, and are related to the Reynolds (\Rey) and Schmidt (\Sc) numbers through $\nu = 1/\rm{R},\ D=\nu/\rm{S}$. As detailed in the following, the initial conditions for eq.~\eqref{eq:pde_sf} are specified by three parameters: the number of shears $n_s$, the number of blobs $n_b$, and the shear width $w$. More specifically, the field $u_x$ consists of $n_s$ uniform shears of width $w$ along the $y$ direction modelled through the function $\tanh{\left( \frac{y-y_k}{n_s w}\right)}$. The field $u_y$, instead, is made of sinusoids $\sin{\left( n_b \pi x\right)}\exp{25|y-y_k|^2 /w^2}$ that decay exponentially in the $x$ direction while exhibiting $n_b$ blobs along $y$. A visual representation of the shear and their width can be found in the insets of Fig.~\ref{fig:m_svn_comp}, but we refer the reader to~\cite{ohana2024well} for further details on the initial conditions and data generation.\\

\paragraph{Methods.---}
The core of our work lies in the encoding of the state of the fluid at a given time as MPS. To this end, we employ the solutions from~\cite{burns2020dedalus,ohana2024well}, which are obtained by solving eq.~\eqref{eq:pde_sf} in the frequency domain. We focus on the solutions contained in the training dataset, and we assume standard mapping of the 2D grid to a comb tensor network, which has been studied in similar conditions in Refs.~\cite{tindall2024compressingmultivariatefunctionstree, peddinti2024quantum, ye2022quantum}. For each time slice, we exactly restructure the 2D tracer $s$ into a linear MPS, then compress the singular values to $\epsilon = 10^{-8}$, which we assume to be lossless (see Fig.~\ref{fig:compression}). Once in this compressed form, we apply any measurements to calculate entropy and magic content of the MPS. Note that we take the maximum bipartite entropy $\tilde{S}_{vN}$ for any bipartition of the MPS. A schematic representation of the encoding is shown in Fig~\ref{fig:sketch}. 
Due to the dataset constraints, the maximum grid is $256 \times 512$ which entails a maximum bond dimension of $\chi =256$. While we employ a comb network mapping~\cite{tindall2024compressingmultivariatefunctionstree} to linearize the two-dimensional grid as shown in Fig.~\ref{fig:sketch}, alternative encodings could also be explored. Each choice entails different trade-offs between bond dimension growth, ease of compression, and of computability of the resource measures, but a systematic comparison across encodings is beyond the scope of the present work (see the appendix for a non-exhaustive exploration).\\

\paragraph{Numerical results.---}
We start our analysis by drawing a phase diagram of the entanglement and non-stabilizerness resources for the tracer across different regimes. To help make the data more consistent, we choose the maximum numbers of shears $n_s =4$ and blobs $n_b =4$ in the initial conditions, and evaluate both quantum resources as a function of the shear width for Reynolds and Schmidt numbers in $\lbrace 10^4, 10^5, 5\times 10^5 \rbrace$ and $\lbrace 0.1,0.2,0.5,1,2,5,10\rbrace$ respectively. By looking at the maximum available shear and blobs, we generally find the most complex evolution of the fluid, and thus in some sense show a `worst case' scenario. Specifically, in Fig.~\ref{fig:phases} we show the behavior of the maximum observed von Neumann entropy $\tilde{S}_{vN}$ and $2-$stabilizer Rényi entropy $\tilde{m}_2$ normalized  with respect to their maximum attainable values. In the following, unless otherwise stated, ``normalized resource" refers to a given resource divided by its maximum attainable value for a given system size. It can be noticed that a larger Reynolds number, which is known to lead to flow instabilities, also results in a higher resource content. Furthermore, the shear width marks a boundary between resource-intensive and resource-efficient regimes which seems to be independent of \Rey \ and \Sc. In some cases, the fluid does not develop vortices, which we largely ignore given initial conditions, but can be partially seen in the phase diagram. 
In order to further investigate the apparent similarity between the two quantum resources, we focus on their evolution in time. In particular, we choose the lowest and highest available Reynolds numbers ($10^4$ and $5\times 10^5$) and evaluate $\tilde{S}_{vN}$ and $\tilde{m}_2$ for different values of S and $w$. As can be seen from the panels in Fig.~\ref{fig:m_svn_comp}, the two quantum resources qualitatively track each other in time, indicating comparable computational complexity within their respective resource theories and substantiating the similarity observed in Fig.~\ref{fig:phases}. Furthermore, we can notice that the central and upper panels confirm that smaller values of $w$ result in more costly states. The lower panels indeed show that, even with a significantly large Reynolds number, a larger width trivializes the evolution. The insets in the panels, displaying the shears at the initial and final time of the evolution, offer a better insight into how the resource content relates to a less stable behavior. A similar analysis on the time evolution of the quantum resources for the pressure field is discussed in the appendix, along with the growth in time of the bond dimension for the simulations in Fig.~\ref{fig:m_svn_comp}.\\

\begin{figure*}[t]
\includegraphics[width=\textwidth]{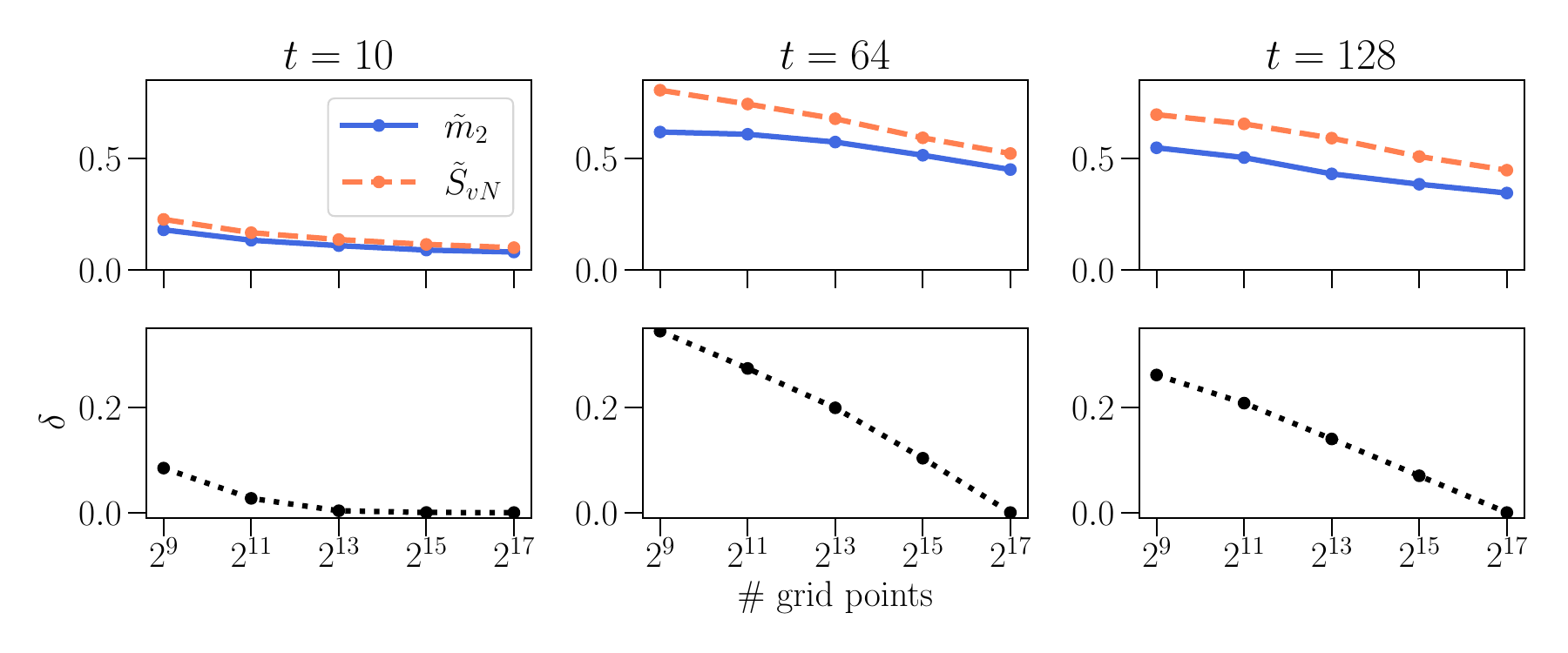}
\caption{\textit{Top panels}: scaling of the normalized non-stabilizerness and entanglement entropy with the number of grid points at early, intermediate and late evolution times. \textit{Bottom panels}: RMSE in the interpolation from the original grid size to the rescaled ones.  All plots refer to the solutions of eq.~\eqref{eq:pde_sf} for the tracer field.}
\label{fig:scaling}
\end{figure*}

\paragraph{Mesh resolution analysis.---}
These results, however, may depend on the refinement of the mesh, that is on the number of grid points or, equivalently, on the number of sites in the MPS encoding. For this reason, we consider the relationship between resources and grid size. We start with the full $(2^{N_x},2^{N_y})$ mesh in the considered dataset, and perform an interpolation to a $(2^{\tilde{N}}_x, 2^{\tilde{N}_y})$ with $\tilde{N}_x < N_x$ and $\tilde{N}_y < N_y$. We use a cubic b-spline interpolation, and study three stages --- early ($t=10$), intermediate ($t=64$), and late ($t=128$) --- in the time evolution of simulation $26$ with $\rm{R}=5\times 10^5$ and $\rm{S}=10$ (top right panel in Fig.~\ref{fig:phases}). We notice that refining the grid, which is essential to capture the detailed behavior of the flowing shears, also plays a non-trivial role in the actual quantum resource content of the encoded MPS. As can be seen from the top panels in Fig.~\ref{fig:scaling}, a too coarse mesh, which corresponds to a less precise representation of the fluid, also results in increased values of both entanglement and non-stabilizerness, i.e. to a more resource-expensive TN or stabilizer based simulation. In the case of entanglement, the increase can be understood as coarser grids artificially introduce long-range correlations between distant points, while finer grids better resolve local structures, allowing a more efficient compression.
We analyze the interpolation error by measuring the root mean squared error (RMSE) defined as
\begin{equation}
    \delta = \sqrt{\frac{1}{N_x N_y} \sum_{i,j} \left( A_{i,j}-\tilde{A}_{i,j}\right)^2}
\end{equation}
where $\tilde{A}_{i,j}$ is obtained by reinterpolating the coarse mesh of size $(2^{\tilde{N}_x}, 2^{\tilde{N}_y})$ back to its original size (bottom panels in Fig~\ref{fig:scaling}).\\

\begin{figure}
\includegraphics[width=0.5\textwidth]{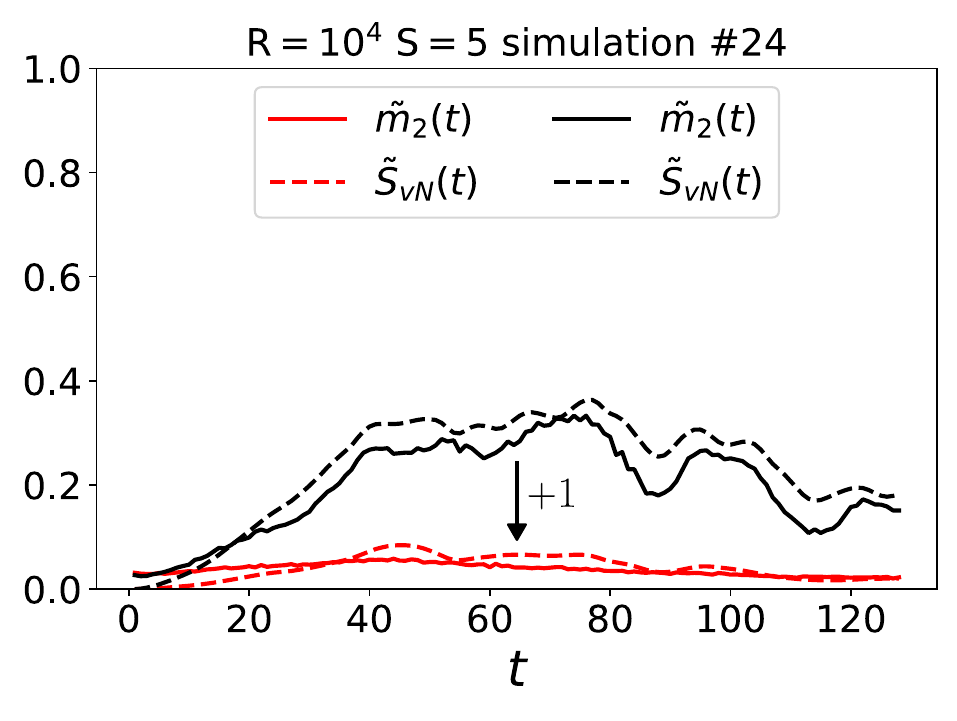}
\caption{Entanglement entropy and non-stabilizerness for non-positive (black lines) and positive (red lines) data. Shifting the values of the data to be positive appears to significantly reduce its complexity. The displayed evolution is the same as in the central left panel of Fig.~\ref{fig:m_svn_comp}.}
\label{fig:sign}
\end{figure}

\paragraph{Relationship to the sign problem.---} Recent work~\cite{chen2024sign2, chen2025sign1} has attributed the presence of non-positive numbers to a difficulty in compressibility of tensor networks. This lack of compressibility is thought to be a manifestation of the so-called quantum Monte Carlo (QMC) sign problem in tensor network simulations. We investigate the role negative numbers play in the resources of our data by shifting the data to be all positive, i.e. by taking $s_{ij} \to s_{ij}+1 \geq 0$. We find a modest decrease in the bond dimension and, as can be seen in Fig.~\ref{fig:sign}, the required resources significantly drop for any time. Specifically, for the chosen simulation, we find that over time the average decrease in bond dimension is $\approx 17 \%$, which leads to an average decrease of $\approx 17 \%$ and $\approx 27 \%$ for entanglement and non-stabilizerness, respectively. Although the norm of the shifted MPS might blow up in some cases, this may still imply that there are more efficient representations of the data than a raw conversion (additional results are shown in the appendix). However, extracting the relevant observables from the data under a more efficient compression, as well as a solid understanding of the role of the sign structure in such compression, is left for future work. \\

\paragraph{Conclusion and outlook.---}
In this work, we assessed the quantum resource content of fluid dynamics data describing the shear flow. To this end, we encoded the classical solutions provided in~\cite{ohana2024well} for the fluid tracer fields as MPSs, and evaluated their entanglement and non-stabilizerness. We calculated a maximum resource phase diagram across different regimes, highlighting how the shear width plays the role of an effective control parameter, marking a boundary between low and high resource cost, independently of fluid parameters. Furthermore, we observed that entanglement and non-stabilizerness follow a similar temporal dynamics for non-trivial evolution, displaying consistent diagnostics of complexity in the encoded state. Our analysis also shows that  factors such as mesh resolution and sign structure crucially influence the quantum resources required for the MPS representation. A too coarse discretization of the domain, besides erasing physical detail, can increase the resource demand of the MPS encoding. Similarly, we found that modifying the sign structure of the data can lead to substantial reductions in both entanglement and non-stabilizerness, suggesting that data preprocessing may be an underexplored lever for achieving more efficient representations. This observation highlights the non-trivial interplay between data structure and quantum resource requirements, emphasizing the potential for optimization in both simulation and compression strategies.

These findings serve as a diagnostic framework for identifying which classical simulations are best suited for efficient representation via tensor networks or stabilizer circuits. The framework could be extended to other systems with a similar structure, or to different TN-encoding of the fluid dynamics data. An important next step is to couple these resource assessments with actual simulation benchmarks on stabilizer- and TN-based solvers, enabling a more complete understanding of hybrid quantum-classical algorithms, and of when and how quantum-inspired methods can outperform classical CFD approaches.\\

\paragraph{Acknowledgments.---}
The authors acknowledge inspiring discussions with Pablo Bermejo. We also thank Rudy Morel for providing insights into the dataset. A.F.M., E.M.S. and R.L. are grateful for the support from the Flatiron Institute, a division of the Simons Foundation.
M. C. acknowledges support from the PNRR MUR project PE0000023-NQSTI, and the PRIN 2022 (2022R35ZBF) - PE2 - ``ManyQLowD''.

\bibliography{bibliography}
\appendix
\section{Appendix A: sign structure of the data}
\begin{figure}[H]
\includegraphics[width=0.5\textwidth]{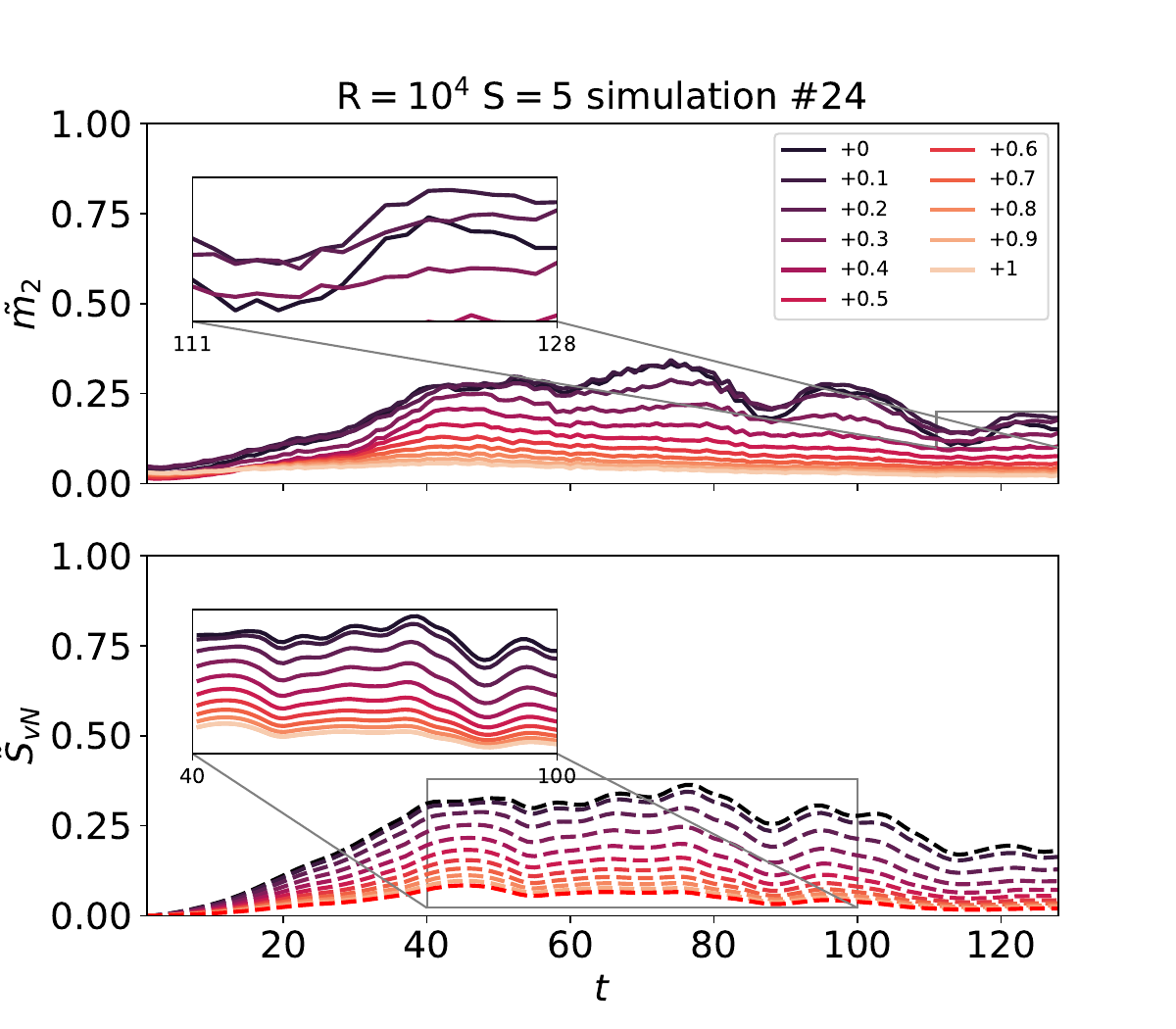}
\caption{Normalized entanglement entropy (bottom panel) and non-stabilizerness (top panel) for data as it is gradually shifted to positive values. While the effect of the sign structure appears to be systematic for entanglement, its role for non-stabilizerness is less structured.}
\label{fig:gradual_shift}
\end{figure}

\begin{figure}[H]
\includegraphics[width=0.5\textwidth]{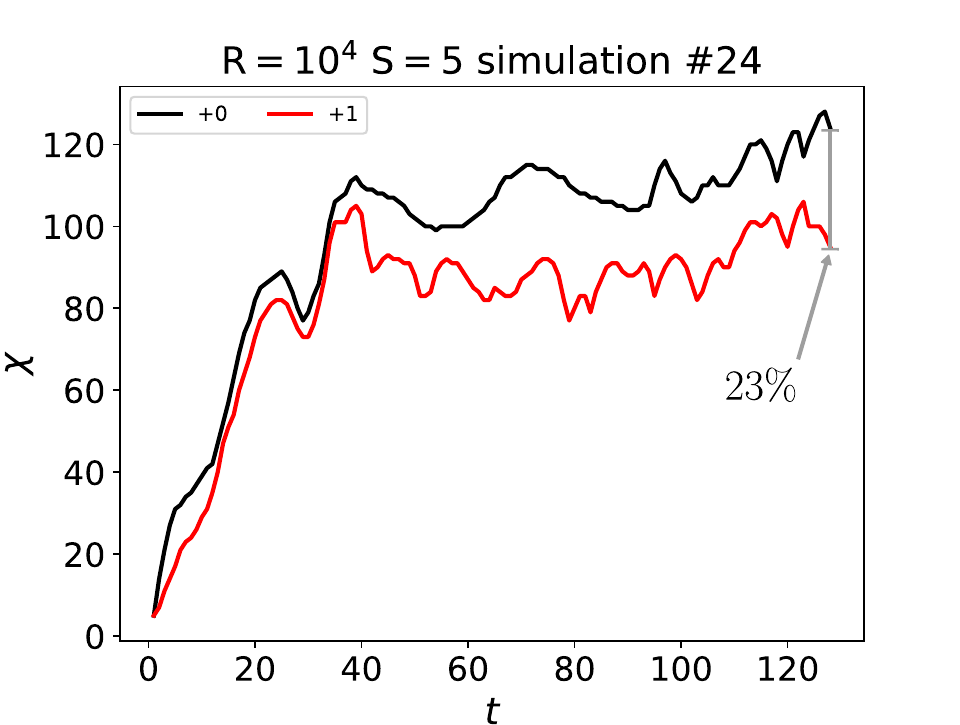}
\caption{Bond dimension $\chi$ of the encoded and truncated MPS as a function of time for the chosen evolution for non-positive (black) and shifted (red) data. The decrease in bond dimension at the latest time is $23\%$.}
\label{fig:chi_evolution}
\end{figure}

\begin{figure*}
\includegraphics[width=\textwidth]{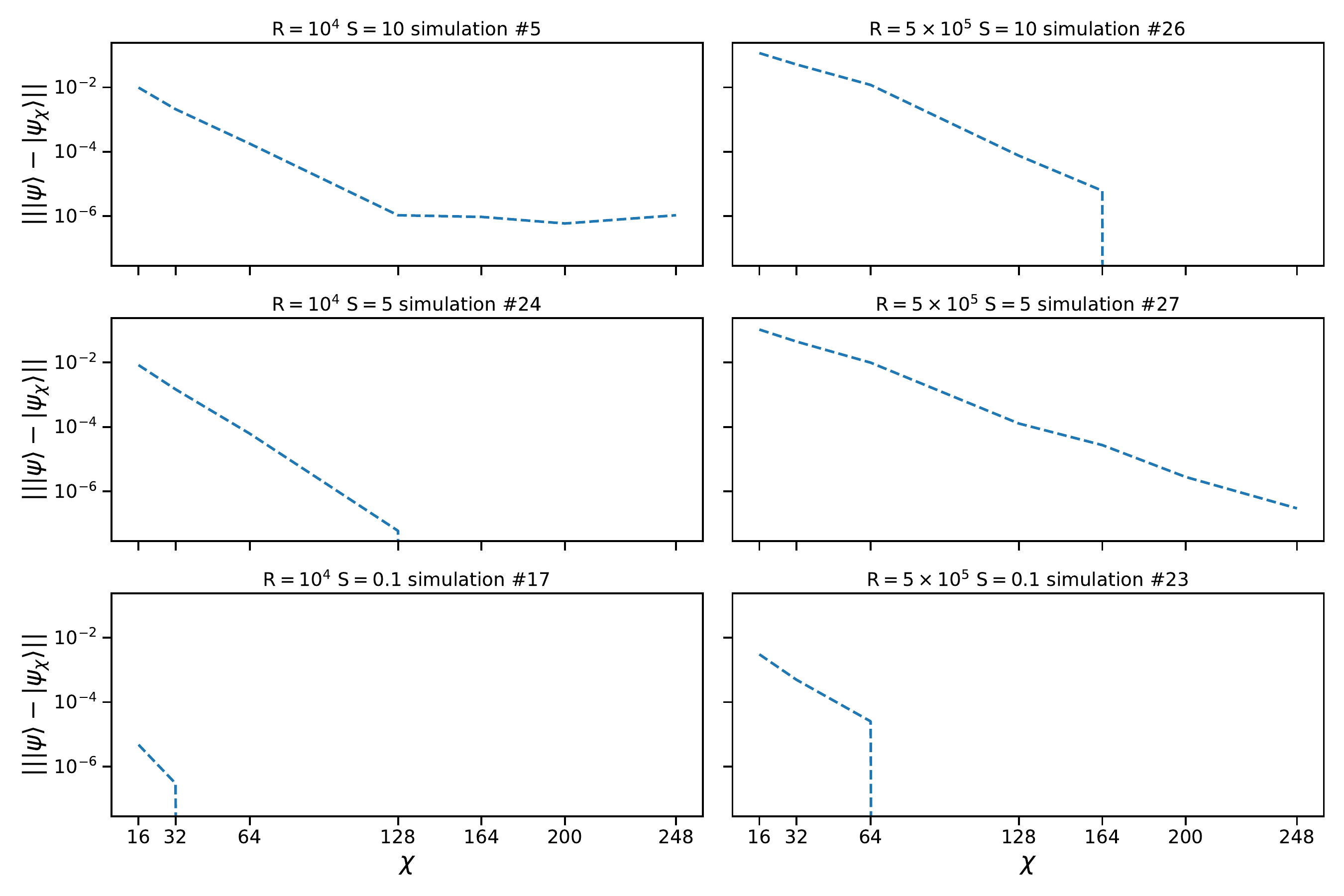}
\caption{Global truncation error as a function of the bond dimension $\chi$. We show it for the final time $t=128$ of the simulations analyzed in Figs.~\ref{fig:m_svn_comp} and~\ref{fig:m_svn_comp_press}. In all the considered cases, the bond dimension achieved by imposing the encoding cutoff ensures the negligibility of such error. The $y-$axis is in log scale.}
\label{fig:all_trunc_err}
\end{figure*}

We present and discuss additional results on the relationship to the sign problem for the data representation. 
To get additional insights about the influence of the sign structure of the data on its quantum resource content, we perform a gradual shift of the original image towards positive values. More specifically, we consider the same simulation analyzed in Fig.~\ref{fig:sign} in the main text, and gradually shift the interval $\left[ -1,1\right]$ to $\left[0, 1\right]$ by transforming the data entries as $s_{ij} \rightarrow s_{ij} + x$, with $x \in \left[ 0,1\right]$ ($x=0$ and $x=1$ corresponding to the original image and to its non-negative version, respectively). Fig.~\ref{fig:gradual_shift} shows that both entanglement and non-stabilizerness are affected by the positivity of the data, but in slightly different ways. In the considered case, entanglement systematically decreases as the data becomes more positive (bottom panel). On the other hand, non-stabilizerness (top panel) appears to only partially follow a similar trend. In fact, it can be seen from the inset that the curves corresponding to $x \leq 0.3$ can be above the $x=0$ one. Fig.~\ref{fig:chi_evolution} shows the effect of the shift on the bond dimension $\chi$ across the evolution. Similarly to the analyzed quantum resources, positive data shows a reduction in $\chi$, which becomes more significant after the first stage of the evolution. As written in the main text, the average decrease in time for $\chi$ is $\approx 17 \%$. 
For completeness, in Fig.~\ref{fig:all_trunc_err} we show the truncation error for the final time $t=128$ of all the simulations in Fig.~\ref{fig:m_svn_comp}.


\section{Appendix B: coarse graining on unstructured data}
\begin{figure}[h!]
\includegraphics[width=0.5\textwidth]{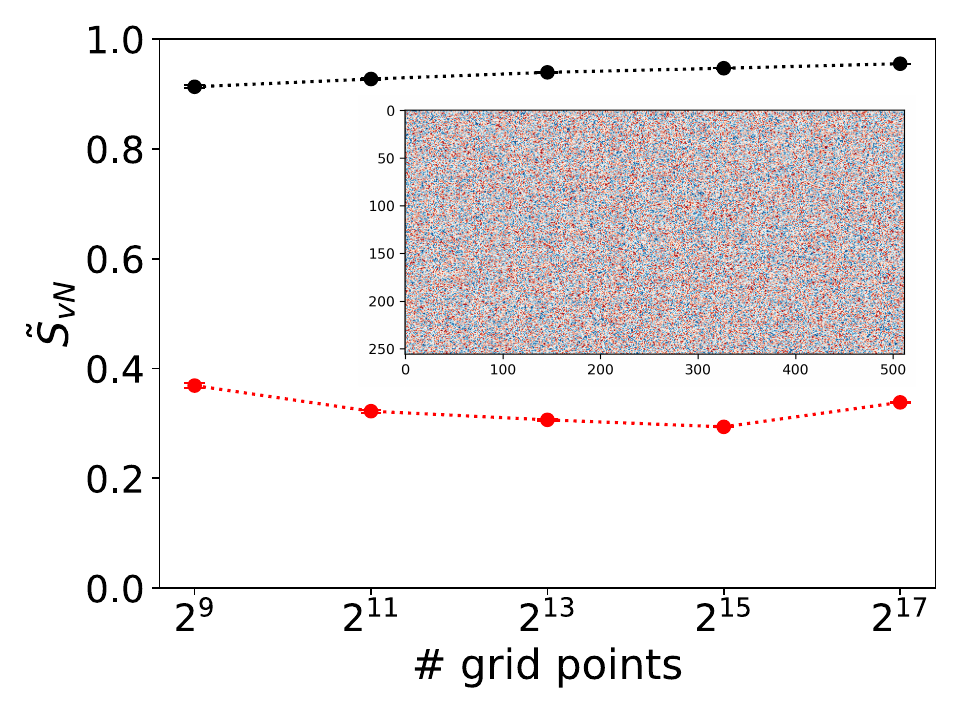}
\caption{Average normalized maximum entanglement entropy for random images with entries uniformly drawn in $\left[ -1, 1\right]$ (black line) and in $\left[0,2\right]$ (red line) after coarse graining them to finer grids. These random images do not show the same behavior as the fluid dynamics data. However, we notice here as well that positive data result in a lower entanglement content. The inset shows an example of the generated matrices.}
\label{fig:random_coarse}
\end{figure}

We further investigate the effect of coarse graining the original grid representing the data on the entanglement entropy of the encoded MPS. For this reason, we initialize $100$ random images as $256 \times 512$ matrices with entries in $\left[ -1,1 \right]$, we interpolate them to smaller grids, and evaluate the normalized maximum entanglement entropy of the resulting MPS. As can be seen from the black curve in Fig.~\ref{fig:random_coarse}, for generic unstructured images, the trend of the entropy is the opposite of the one observed for physical data. Indeed, the smaller the grid the less the entanglement content. In addition, we perform the same analysis on positive random images with entries in $\left[ 0,2 \right]$. The red curve in Fig.~\ref{fig:random_coarse} displays a significant decrease in the entanglement content, which is notably not linked to a lower bond dimension MPS encoding. 

\begin{figure*}
\includegraphics[width=\textwidth]{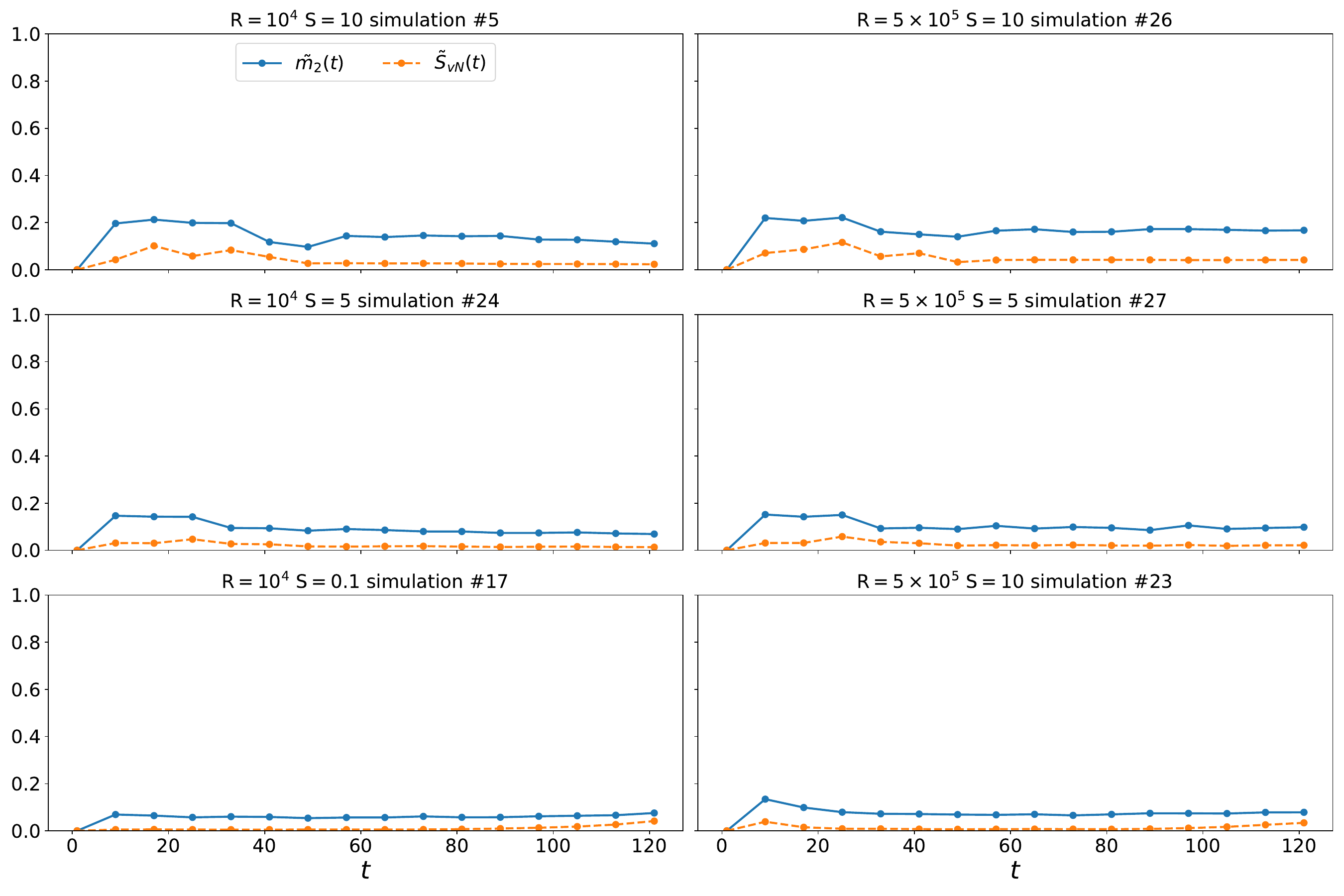}
\caption{Comparison of the normalized non-stabilizerness $\tilde{m}_2(t)$ and normalized entanglement entropy $\tilde{S}_{vN}(t)$ for the pressure field as a function of time for different values of \Rey, \Sc, and initial shear width $w \in [1,3,4]$.  }
\label{fig:m_svn_comp_press}
\end{figure*}

\section{Appendix C: resource content for the pressure field}
We show some results for the entanglement and non-stabilizerness content of the pressure field. Specifically, we perform the same analysis as in Fig.~\ref{fig:m_svn_comp}, assessing the evolution of the two resources at maximum amount of blobs and shear. In analogy with the tracer, from Fig.~\ref{fig:m_svn_comp_press} we see that the two resources track each other in time, and that a smaller shear width increases the complexity of the evolution. However, it is worth noticing that their normalized values are significantly lower, and that for the pressure field the non-stabilizerness becomes the dominant resource. In this case, due to the triviality of the evolution, fewer time points were taken into account.

\section{Appendix D: faithfulness of the positive data encoding}
\begin{figure*}
\includegraphics[width=\textwidth]{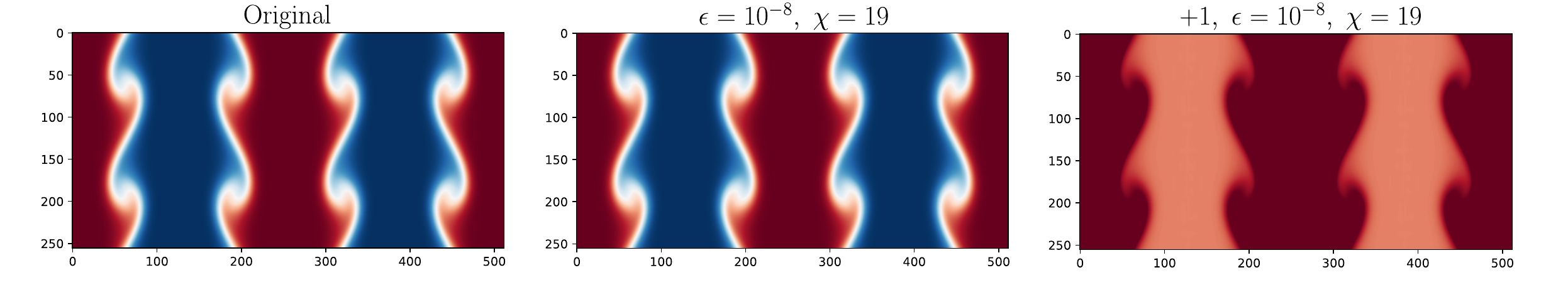}
\caption{ \textit{Left panel}: original data of Fig.~\ref{fig:sketch}. \textit{Central panel}: lossless representation of the tracer field data obtained from shifted values encoded as MPS. \textit{Right panel}: visual representation of the shifted data.}
\label{fig:positive_compression}
\end{figure*}

We provide evidence for the fact that shifting the data to positive values preserves its structure, and allows to easily retrieve the non-positive encoding. In Fig.~\ref{fig:positive_compression}, we show the original image (left panel) and the one obtained from the positive MPS (right panel). The central panel, instead, is obtained from the positive encoded MPS as follows. After shifting the data and compressing it into an MPS, we contracted the MPS auxiliary dimensions and reshaped it as an two-dimensional array. Unshifting the latter, we obtained what is shown in the panel, which is a lossless representation of the data.

\section{Appendix E: encoding order}
\begin{figure}[H]
\includegraphics[width=0.5\textwidth]{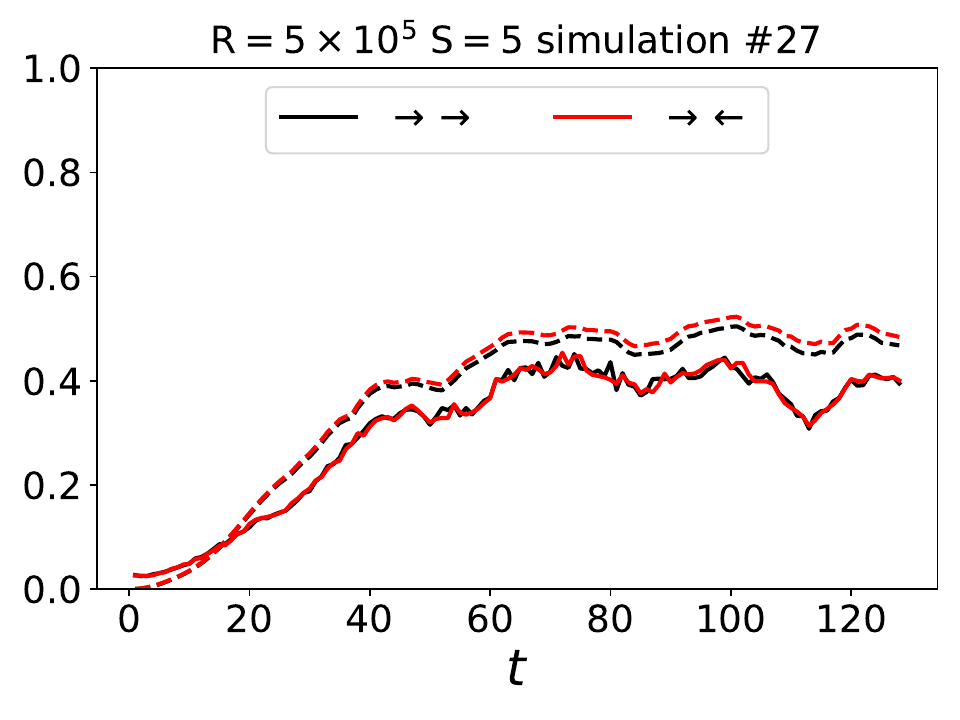}
\caption{Normalized entanglement (dashed lines) and non-stabilizerness (solid lines) for different MPS ordering of the encoded state. For the specific analyzed simulation, the encoding order does not change the resource content of the MPS.}
\label{fig:ordering_comparison}
\end{figure}

The sequence in which the grid's elements are arranged for the MPS encoding can significantly impact the compressibility of the state, and the computational resources required to describe it. Although a comprehensive exploration of all possible orderings is a complex task, in this appendix we compare the ordering chosen throughout the main text with an alternative that has shown better performance in some scenarios~\cite{ye2022quantum}. Specifically, we consider the evolution in the central right panel in Fig.~\ref{fig:m_svn_comp}, and compare the encoding in Fig.~\ref{fig:sketch} with the one where the $y$ entries are in reverse order. We refer to these encodings as $\longrightarrow \longrightarrow$ and $\longrightarrow \longleftarrow$, respectively. Fig.~\ref{fig:ordering_comparison} shows the normalized entanglement (dashed lines) and non-stabilizerness (solid lines) for both encodings. It can be noticed that, for this specific and arguably non trivial evolution, the order does not remarkably affect the resource content of the state, presumably due to the periodic structure of the shear flow.



\end{document}